\documentclass{article}
\usepackage{spconf,amsmath,graphicx}
\usepackage{subfigure}
\usepackage{booktabs} 
\usepackage{array}
\usepackage{hyperref}
\usepackage{marvosym}

\title{MelGlow: Efficient Waveform Generative Network Based on \\Location-Variable Convolution}
\name{Zhen Zeng, Jianzong Wang$^*$\thanks{$^*$Corresponding author: Jianzong Wang, jzwang@188.com}, Ning Cheng, Jing Xiao}
\address{Ping An Technology (Shenzhen) Co., Ltd.}
\begin{document}
%
\maketitle
\begin{abstract}
  Recent neural vocoders usually use a WaveNet-like network to capture the 
  long-term dependencies of the waveform, but a large number of parameters 
  are required to obtain good modeling capabilities. 
  In this paper, an efficient network, named location-variable convolution, 
  is proposed to model the dependencies of waveforms. 
  Different from the use of unified convolution kernels in WaveNet to 
  capture the dependencies of arbitrary waveforms, location-variable convolutions 
  utilizes a kernel predictor to generate multiple sets of convolution 
  kernels based on the mel-spectrum, where each set of convolution kernels 
  is used to perform convolution operations on the associated waveform intervals. 
  Combining WaveGlow and location-variable convolutions, 
  an efficient vocoder, named MelGlow, is designed. 
  Experiments on the LJSpeech dataset show that MelGlow achieves 
  better performance than WaveGlow at small model sizes, 
  which verifies the effectiveness and potential optimization
  space of location-variable convolutions.

\end{abstract}
\begin{keywords}
  speech synthesis, text-to-speech, neural vocoder, glow, location-variable convolution
\end{keywords}

\section{Introduction}

Recently, speech synthesis is playing an increasingly important role in 
the field of human-computer interaction.  
Since small changes in voice 
quality or latency can have a large impact on customer experience, 
it is necessary for speech product design to efficiently synthesize
high quality speech, especially for smart speakers and dialogue robots. 
Currently, the general speech synthesis pipeline is composed of two components 
\cite{Char2wav,Tacotron,Tacotron2,TransformerTTS,DeepVoice3,FastSpeech,ParaNet,real-time-tts,ProsodyLearn}: 
(1) a speech feature prediction model that transforms the text into time-aligned 
speech features, for example mel-spectrum, and (2) a neural vocoder that generates 
raw waveform from speech features. Our work focuses on the second model to 
efficiently generate speech with high quality from the mel-spectrum.

Traditional speech vocoders use digital signal processing (DSP) to reconstruct 
speech waveforms at very high rate, but their quality has always been  
limited \cite{LinearPredictionVocoder1,LinearPredictionVocoder2,GriffinLim,WorldVocoder}. 
In recent research, neural vocoders have been widely studied. 
WaveNet \cite{wavenet}, an autoregressive generative model, is proposed to synthesize 
high-quality speech at a high temporal resolution. 
WaveRNN \cite{WaveRNN} applies an efficient recurrent 
neural network to generate speech at 4 times faster than real-time and 
uses a weight pruning technique to make it deployable on mobile CPU. 
LPCNet \cite{LPCNet} combines WaveRNN with linear prediction to significantly improve the efficiency of
speech synthesis. 
Parallel WaveNet \cite{ParallelWavenet} and ClariNet \cite{Clarinet} use Inverse Autoregressive 
Flows (IAF) \cite{IAF} as the student model to distill knowledge from an autoregressive 
teacher WaveNet. 
FloWaveNet \cite{FloWaveNet} and WaveGlow \cite{WaveGlow} design a flow-based network \cite{Flow} 
to generate high-quality speech and have a very simple training process. 
In \cite{GAN-Excited-Linear-Prediction}, a GAN-excited linear prediction is presented to generate waveform, 
where the GAN generator outputs the entire excitation signal of the linear prediction.
MelGAN \cite{MelGAN} and GAN-TTS \cite{GAN-TTS} implement a Generative Adversarial Network (GAN) \cite{GAN}
to directly generate waveforms. Similarly, 
based on the GAN, Parallel WaveGAN \cite{Parallel-WaveGan} proposes a joint optimization method 
between the adversarial loss and multi-resolution STFT loss \cite{STFTLoss} to 
capture the time-frequency distributions of the realistic speech signal. 
These GAN-based vocoders is highly parallelisable due to the design of the efficient feed-forward generator.

The key of neural vocoder is how to model the long-term dependence of waveform 
based on the conditioning acoustic features. 
In WaveNet \cite{wavenet}, the dilated causal convolution is used to capture 
the long-time dependencies of waveform. 
where the mel-spectrum is used as the local conditioning input and 
added into the gated activation unit. 
In \cite{pitch-convolution}, a pitch dependent dilated convolution is designed 
to improve the pitch controllability of WaveNet.
Due to the superior performance of the WaveNet, 
the dilated convolution structure is used by many subsequent neural vocoders 
\cite{Parallel-WaveGan,GAN-TTS,ShallowWavenet,WaveFFJORD}.
Although the dilated convolution enables the model to 
capture the long-term dependence of the waveform, 
a large number of convolution kernels are required in each layer 
of the dilated convolution to capture different time-dependent 
features for achieving good performance.

In this work, we propose the location-variable convolution 
for the design of the waveform generation network
in order to capture the long-term dependency more efficiently. 
In detail, a kernel predictor is designed to predict the coefficients 
of the convolution kernels according to the conditioning acoustic features (e.g. mel-spectrum), 
and different segments of the waveform use the different convolution kernels 
to implement the convolution operation. This convolution method is used to 
redesign the network structure of WaveGlow \cite{WaveGlow} so as to achieve a more efficient 
speech vocoder, named MelGlow. Experiments on the LJSpeech dataset \cite{ljspeech17} show that 
higher quality speech is acquired by MelGlow in the case of  a limited number of model parameters. 
Meanwhile, as the mode size decreases, the rate of performance degradation 
of MelGlow is significantly slower than that of WaveGlow.
And the main contributions of our works as follow:
\begin{itemize}
  \item A novel convolution method, named location-variable convolution, 
  is proposed to efficiently capture the long-term dependency, 
  where the coefficients of the convolution kernels are predicted 
  by a kernel predictor according to the conditioning input;

  \item A more efficient speech vocoder, named MelGlow, 
  is designed based on location-variable convolutions, 
  and shown to generate higher quality speech than 
  WaveGlow when limiting the model size;

  \item The relationship between the performance of MelGlow and the number 
  of parameters is analyzed experimentally, which illustrates that MelGlow
  has great potential for optimization in the future. 
\end{itemize}

\section{Location-Variable Convolution} 

In this section, we propose the location-variable convolution 
to model the long-term dependency of input sequence 
based on the local conditioning sequence, 
where the coefficients of the convolution kernels 
change throughout the convolution process. 

\subsection{Motivation}
Reviewing the traditional vocoder using digital signal 
processing (DSP) method to reconstruct speech waveforms \cite{LinearPredictionVocoder1, LinearPredictionVocoder2}, 
one popular method is the linear prediction vocoder \cite{LinearPredictionVocoder1} which 
uses the Levinson-Durbin algorithm to calculate the linear 
prediction coefficients of a simple all-pole linear filter \cite{LinearPrediction}
according to acoustic features. The prediction process of 
the linear predictor is similar to the convolution process 
of the causal convolution, except that the coefficients of 
the linear predictor are calculated based on the acoustic 
features at different frames and the coefficients of convolution kernels 
in the causal convolution is the same in the whole process. 
Inspired by this difference, we design a novel convolution with 
variable convolution kernel coefficients to model the long-time 
dependencies of waveforms more efficiently.

\begin{figure}[t]
  \centering
  \includegraphics[width=0.7\linewidth]{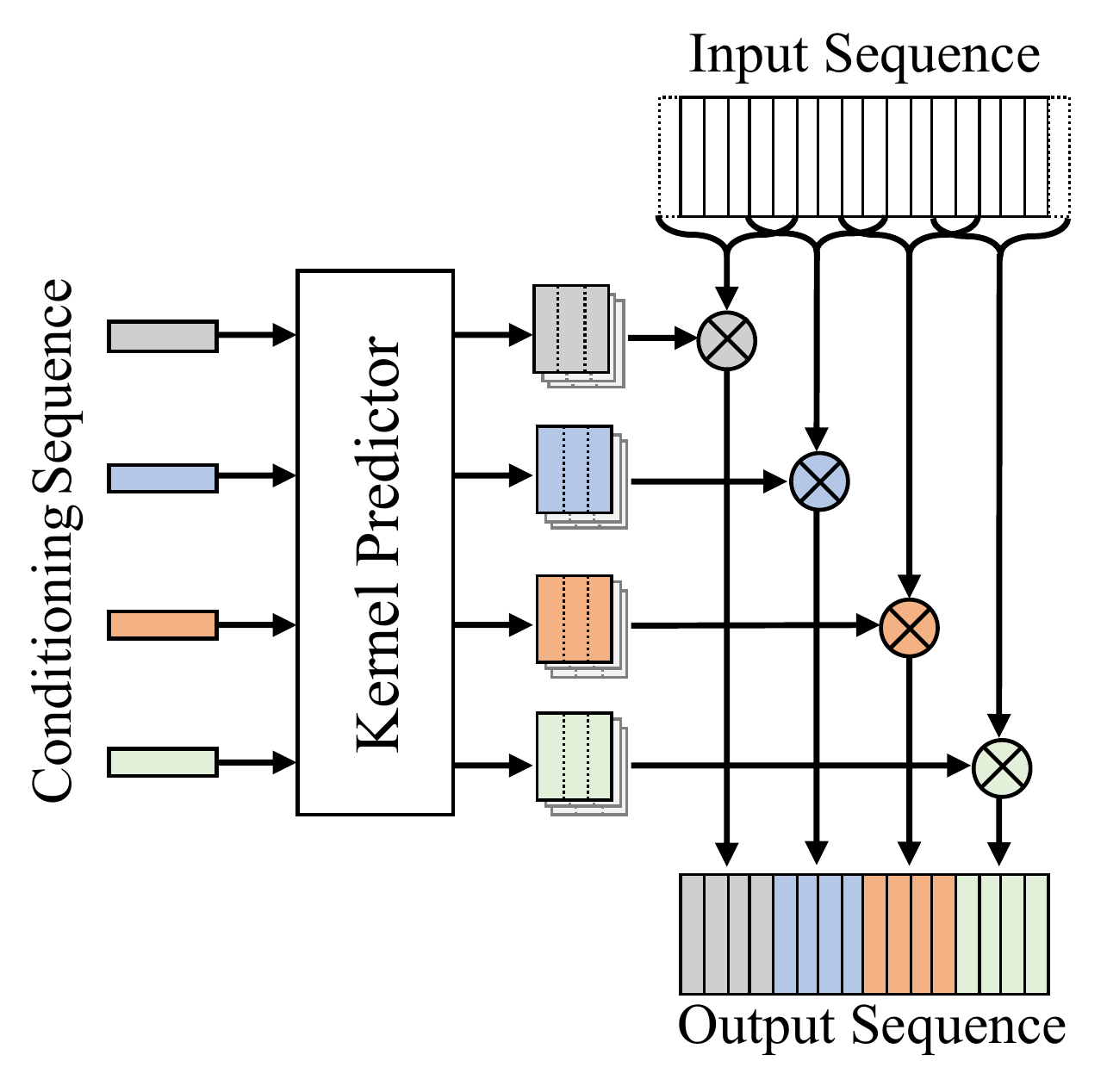}
  \caption{An example of convolution process in the location-variable convolution. 
  According to the conditioning sequence, the kernel predictor generates multiple sets of 
  convolution kernels, which are used to perform convolution operations on 
  the associated intervals in the input sequence. 
  Each element in the conditioning sequence corresponds to 4 elements in the input sequence. 
  }
  \label{figure-lvc}
\end{figure}

\subsection{Structure}

Define the input sequence to the convolution as 
$\boldsymbol{x}=\{x_1,x_2,$ $ \ldots, x_n\}$, and define the local conditioning 
sequence as $\boldsymbol{h}=\{h_1,h_2,\ldots,h_m\}$. An element in the local conditioning sequence is 
associated with a continuous interval in the input sequence. 
In order to model the feature of the input sequence, 
the location-variable convolution uses a novel convolution method, 
where different intervals in the input sequence use 
different convolution kernels to implement the convolution operation. 
In detail, a kernel predictor is designed to predict multiple 
sets of convolution kernels according to the local conditioning sequence. 
Each element in the local conditioning sequence corresponds to a set 
of convolution kernels, which is used to perform convolution operations on 
the associated intervals in the input sequence. 
And the output sequence is spliced by the convolution results on 
each interval. 

Similar to WaveNet, the gated activation unit is also applied, 
and the local condition convolution can be expressed as
\begin{align}
  & \{ \boldsymbol{x}_{(i)} \}_m = \text{split}( \boldsymbol{x} )  \\
  \{ \boldsymbol{W}^f_{(i)}, & \boldsymbol{W}^g_{(i)} \}_m = \text{Kernel Predictor}( \boldsymbol{h} ) \\
  \boldsymbol{z}_{(i)} &= \tanh ( \boldsymbol{W}^f_{(i)} * \boldsymbol{x}_{(i)} ) \odot \sigma ( \boldsymbol{W}^g_{(i)} * \boldsymbol{x}_{(i)} ) \\
  \boldsymbol{z} &= \text{concat} ( \boldsymbol{z}_{(i)} )
\end{align}
where $\boldsymbol{x}_{(i)}$ denotes the intervals of the input sequence associated with $ h_i $, 
$\boldsymbol{W}^f_{(i)}$ and $\boldsymbol{W}^g_{(i)}$ denote the filter and gate convolution kernels 
for $\boldsymbol{x}_{(i)}$. 

\begin{figure*}[t]
  \subfigure[ Architecture of MelGlow ]{
    \label{figure-melglow}
    \begin{minipage}[b]{0.3\linewidth} 
        \centering 
        \includegraphics[width=0.56\linewidth]{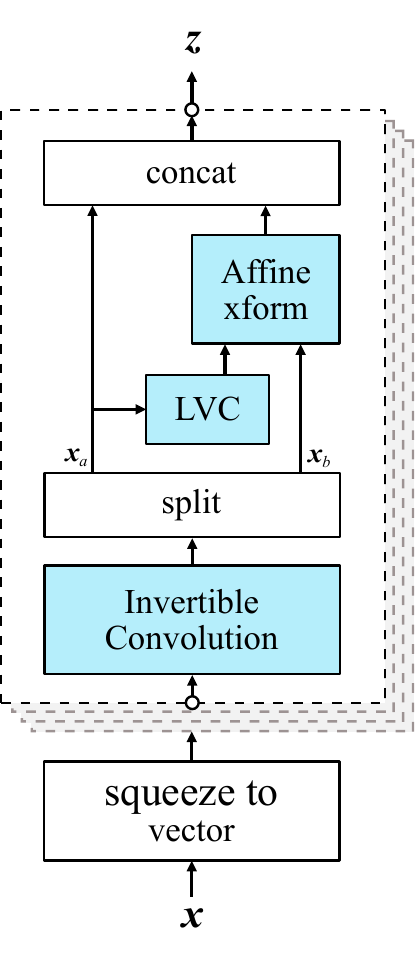}
    \end{minipage}
  }
  \subfigure[ Location-Variable Convolutions of MelGlow ]{
    \label{figure-lvc-in-melflow}
    \begin{minipage}[b]{0.38\linewidth} 
        \centering 
        \includegraphics[width=0.80\linewidth]{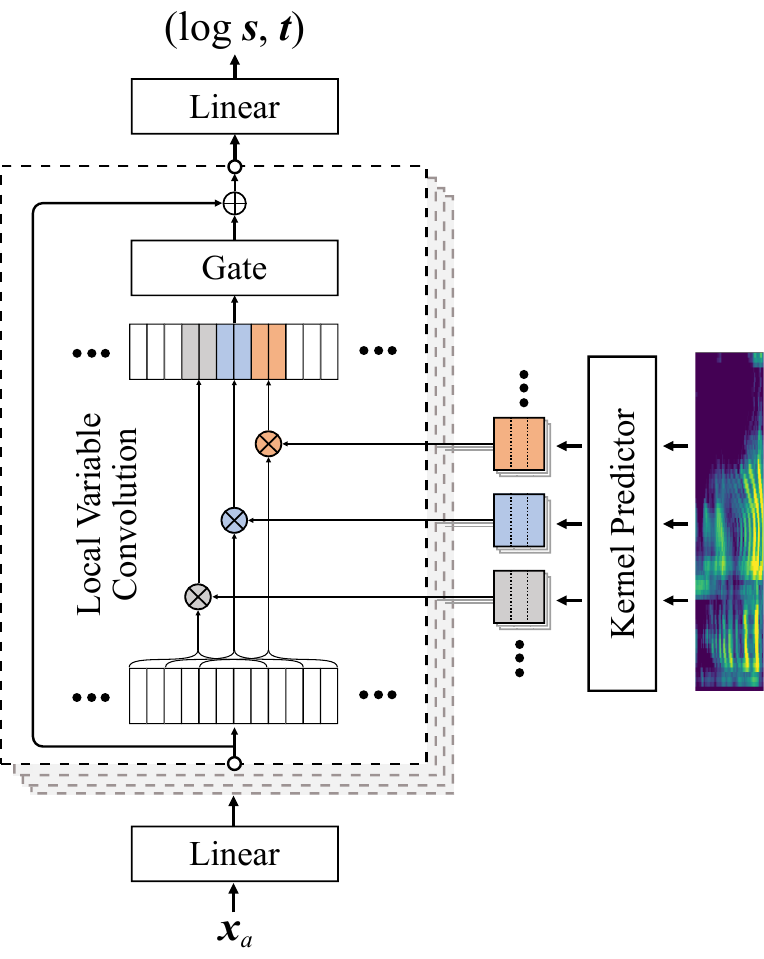}
    \end{minipage}
  }
  \subfigure[ Kernel Predictor ]{
    \label{figure-kernel-predictor}
    \begin{minipage}[b]{0.3\linewidth} 
        \centering 
        \includegraphics[width=0.56\linewidth]{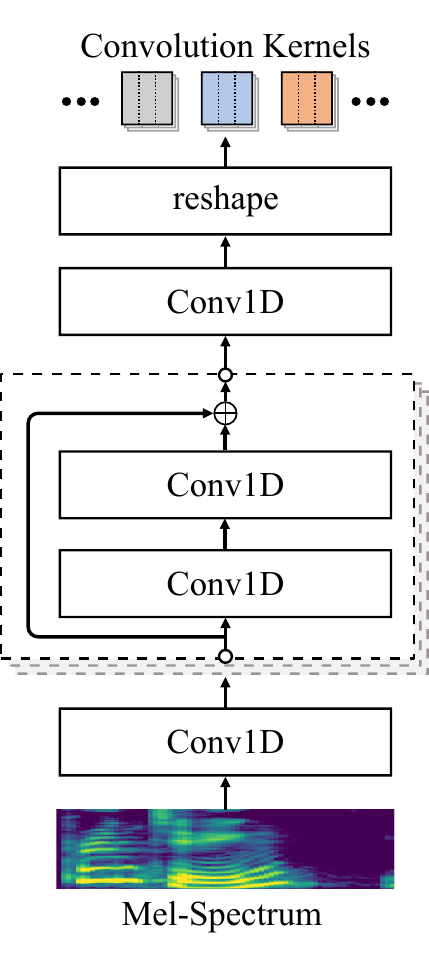}
    \end{minipage}
  }
  \caption{ (a) The architecture of MelGlow, which is similar to WaveGlow, except that 
  the WN block is replaced by location-variable convolutions (LVC).
  (b) The process of location-variable convolutions in MelGlow, 
  which is applied according to the correlation between mel-spectrums and waveform.
  (c) The structure of the kernel predictor in MelGlow, 
  which consists of multiple 1D convolution layers. }
\end{figure*}

For a more visual explanation, Figure \ref{figure-lvc} shows an example of the location-variable convolution, 
where the input sequence contains 16 elements and the conditioning sequence contains 4 elements. 
Each element in the conditioning sequence corresponds to 4 elements in the input sequence.
Taking the conditioning sequence as input, 
the kernel predictor generates multiple convolution kernels, 
each of which is used to perform 1D convolution operation on 4 elements of the input sequence 
with one padding.

Note that, since location-variable convolutions can generate different kernels for different 
conditioning sequences, it has more powerful capability of modeling the long-term dependency than 
traditional convolutional network. However, the key is how to design the structure of 
the kernel predictor, which has a significant impact on model performance.

\section{MelGlow} 

In this section, location-variable convolutions is used to 
design an efficient neural vocoder based on WaveGlow, 
which has fewer model parameters, but matches WaveGlow 
in term of speech quality.

\subsection{Architecture}

As shown Figure \ref{figure-melglow}, we keep the main architecture of WaveGlow 
and use location-variable convolutions instead of the WN block. 
To make it easier to distinguish, the novel vocoder is called as MelGlow. 
The waveform data is firstly squeezed to a sequence with 8 channels, 
which is passed through multiple flow layers to output a latent variable. 
In each flow layer, the input sequence is mixed across channels 
by an invertible $1 \times 1$ convolution, and then is split into 
two halves $(\boldsymbol{x}_a, \boldsymbol{x}_b)$ along the channel dimension. 
$ \boldsymbol{x}_a $ is passed through location-variable convolutions with 
the mel-spectrum as the conditioning sequence to produce multiplicative 
terms $ \boldsymbol{s} $ and additive terms $ \boldsymbol{b} $. 
$ \boldsymbol{x}_b $ is scaled and translated by these terms, 
and then is concatenated with $ \boldsymbol{x}_a $ as the output sequence. 
The output sequence of the last flow layer is used to compute 
the log-likehood of the spherical Gaussian, which is added to the log Jacobian 
determinant of the invertible convolution as the final objective function, 
given by the log multiplicative term $ \boldsymbol{s} $.

\subsection{Location-Variable Convolution for MelGlow}

Since the mel-spectrum is calculated from the waveform using short-time 
Fourier Transform (STFT) and a series of transformation operations, 
each element of the mel-spectrum sequence is related to a contiguous  
segment of samples in the waveform. Therefore, the location-variable convolution 
is applied based on this correlation between the mel-spectrum and waveform.

The location-variable convolution uses $ \boldsymbol{x}_a $ as the input sequence, and 
the mel-spectrum as the local conditioning sequence. 
The mel-spectrum is passed into the kernel predictor to predict multiple 
sets of convolution kernels, and each set of convolution kernels is used 
to perform convolution operations on the associated intervals of the input 
sequence, whose association is based on the calculation process of STFT.
For example, if the Hann window size is 1024 and the hop length is 256 
in STFT, each element of the mel-spectrum sequence is associated with 
1024 waveform samples with 256 offsets for the next element. 
Due to the waveform being squeezed to a vector with 8 channels, each set of 
convolution kernels is used to perform dilated convolution operation on 
128 elements of the input sequence with 32 offsets for next set. 
In addition, multiple local condition convolution layers with different dilation 
are stacked to expand the receptive fields, as shown in Figure \ref{figure-lvc-in-melflow}.

\subsection{Kernel Predictor}

The kernel predictor is used to predict the convolution kernels according to 
the mel-spectrum. In our case, the kernel predictor is composed of multiple 1D 
convolution layers and a linear layer, as shown in Figure \ref{figure-kernel-predictor}. 
The kernel size of the first 1D convolution layer is set to two 
without padding to adjust the alignment between the mel-spectrum sequence 
and the input sequence. 
For example, if the Hann window size is 1024 and the hop length 
is 256 in STFT, the waveform will be padded with 512 samples 
at the start and at the end, and the length of the mel-spectrum 
sequence is one greater than the length of the waveform divided by the hop length. 

Followed by the first 1D convolution layer, multiple residual convolutional 
layers is applied, where each residual convolutional layer contains a 2-layer 
1D convolution. 
The tanh activation function and the batch normalization \cite{BatchNormalization} 
is used in each convolution layer. 
The linear layer is used to change the number of 
the channels and output the coefficients of the convolution kernels for 
location-variable convolutions, where the number of the output channels 
is determined by the number of the kernel coefficients in the location-variable 
convolution. Lastly, the output of the linear layer is reshaped into the form 
of the convolution kernel, which is used to perform location-variable convolutions 
on the input sequence.


\section{Experiments} 

\subsection{Training Details}

\subsubsection{Dataset}

We train our MelGlow model on the LJSpeech dataset \cite{ljspeech17}. 
This dataset contains 13,100 English audio clips 
of a single female speaker with 
a total duration of approximately 24 hours. 
The entire dataset is randomly divided into 3 sets: 
12,600 utterances for training, 400 utterances for validation, and 100 utterances for test. 
The sample rate of audio is set to 22,050 Hz. 
The mel-spectrograms are computed through a short-time 
Fourier transform with Hann windowing, 
where 1024 for FFT size, 1024 for window size 
and 256 for hop size. The STFT magnitude is 
transformed to the mel scale using 80 channel 
mel filter bank spanning 60 Hz to 7.6 kHz.

\subsubsection{Model Configuration} 

In order to analyze the efficiency of the location-variable convolution, 
we set the configuration of MelGlow to be the same as WaveGlow on 
the main architecture. In detail, MelGlow has 12 flow layers 
and output 2 of the channels after every 4 flow layers. 
Each flow layer has 7 layers of location-variable convolutions 
with the same dilated convolution method as WaveGlow. 
The hidden channel of the kernel predictor is set to 64, 
and the number of the residual convolution layer is set to three. 
In addition, the channel in location-variable convolutions 
is set to 128, 64, 48 and 32 for comparison experiments.

We train MelGlow on a NVIDIA V100 GPU, with batch size of 8 samples. 
Each sample is a one-second audio clip randomly selected from the 
training dataset. We used the Adam optimizer \cite{Adam} with a learning rate of 
$10^{-4}$ and 600,000 iterations to train it. When training appeared to plateau, 
we adjusted the learning rate to $5\times10^{-5}$. 
As comparative experiments, WaveGlow with different channel number (512, 256, 128 and 64) 
in the dilated convolution layers is chosen as baseline model, 
which is trained following \cite{WaveGlow}.

\subsection{Results} 

\subsubsection{Vocoder Evaluation}

In order to evaluate the performance of MelGlow, 
we use the mel-spectrograms extracted from test utterances 
as input to obtain synthetic audio, which is rated together with 
the ground truth audio (GT) by 50 native English speakers with headphones 
in a conventional mean opinion score (MOS) evaluation.
evaluation indicator of MelGlow. At the same time, 
the audios generated by Griffin-Lim algorithm \cite{GriffinLim} and WaveGlow \cite{WaveGlow} are also rated together. 

The evaluation results are shown in Table \ref{Table-Vocoder}, where the number of model 
parameters also is illustrated.  
Although the best performance of MelGlow in our case does not outperform WaveGlow, 
MelGlow can obtain a better quality of speech at small model size. 
Meanwhile, as the mode size decreases, 
the rate of performance degradation of MelGlow is significantly 
slower than that of WaveGlow. 
According to our analysis, 
in the case of limited model size, MelGlow can still generate a variety 
of convolution kernels to capture the long-term dependent features, 
which enables the model to maintain good performance.

\begin{table}[t]
  \begin{center}
  \caption{The comparison of MOS with 95\% confidence intervals between MelGlow and WaveGlow.}
  \begin{tabular}{p{2.8cm}p{2cm}<{\centering}p{2cm}<{\centering}}
  \toprule
  \textbf{Method}&\textbf{Parameters}&\textbf{MOS} \\
  \midrule
  GT & $-$ & $4.56 \pm 0.05$  \\
  Griffin-Lim  & $ - $         &   $ 3.28 \pm 0.24 $  \\
  WaveGlow-64  & $  17.59 $ M  &   $ 3.65 \pm 0.15 $  \\
  WaveGlow-128 & $  34.83 $ M  &   $ 4.12 \pm 0.13 $  \\
  WaveGlow-256 & $  87.87 $ M  &   $ 4.29 \pm 0.11 $  \\
  WaveGlow-512 & $ 268.29 $ M  &   $ 4.38 \pm 0.08 $  \\
  \midrule
  MelGlow-32   & $ 19.3 $ M   &  $ 4.11 \pm 0.12 $  \\
  MelGlow-48   & $ 40.4 $ M   &  $ 4.16 \pm 0.09 $  \\
  MelGlow-64   & $ 71.3 $ M   &  $ 4.18 \pm 0.10 $  \\
  MelGlow-128  & $ 294.3 $ M  &  $ 4.19 \pm 0.08 $  \\
  \bottomrule
  \end{tabular}
  \label{Table-Vocoder}
  \end{center}
\end{table}

\subsubsection{Text-to-Speech Evaluation} 

Since speech vocoders are a component in text-to-speech systems, 
the vocoder and the mel-spectrum prediction network can be tested 
together to better evaluate the performance. In our experiments, 
Tacotron2 (T2) \cite{Tacotron2}, FastSpeech (FS) \cite{FastSpeech} and 
AlignTTS (AT) \cite{AlignTTS} are used as the mel-spectrum prediction model, 
which predict mel-spectrograms based on the input text sequence. 
MelGlow with 32 channels and WaveGlow with 128 channels are used to 
generate waveforms based on the predicted mel-spectrums, 
which are rated together. 
As shown in Table \ref{Table-TTS}, synthesized audios from MelGlow have similar quality as
that from WaveGlow. This indicates that MelGlow can 
achieve performance that matches WaveGlow, but has fewer model parameters 
when limiting the model size.

\subsection{Analysis of Kernel Predictor}

Since MelGlow uses the kernel predictor to generate various 
convolution kernels for convolution operations, 
the design of the kernel predictor has a significant impact 
on the performance. We therefore analyzed the impact of the number of 
residual convolution layers and hidden channels in the kernel predictor. 
In particular, we set the number of residual convolution layers to 1, 3, 5 
and the hidden channels to 32, 64 and 128 for comparison experiments.
The evaluation results of these model are shown in Table \ref{Table-Analysis}. 
We can find that models of sufficient size give satisfactory 
performance, and more complicated networks do not achieve a significant 
performance improvement. 
One of the reasons is that 
the coefficients of convolution kernel 
predicted by the kernel predictor is not stable enough,  
and MelGlow is more difficult to converge 
if the network of the kernel predictor has too much parameters. 

In WaveNet, the 1D dilated convolution is adopted to model the dependencies 
of waveform, where the convolution kernel is used as the waveform time feature extraction tool. 
In our opinion, due to the mutual independence of the acoustic features (such as mel-specturms), 
we can use difference convolution kernels to implement convolution operations 
on difference time intervals to obtain more effective feature modeling capabilities.
Our exploratory experiment verified the feasibility of this approach, 
but the variability of the convolution kernel coefficients also affects 
the stability and convergence of the entire model.    
However, we believe that designing a more reasonable network for 
the kernel predictor can make MelGlow obtain better performance, 
and further research on the best results that 
the variable convolution can achieve is also our follow-up task.

\begin{table}[t]
  \begin{center}
    \caption{The comparison of MOS with 95\% confidence intervals and time cost in text-to-speech system.}
  \begin{tabular}{p{3cm}p{1.8cm}<{\centering}p{2cm}<{\centering}}
  \toprule
  \textbf{Method}&\textbf{MOS}&\textbf{Time Cost} (s) \\
  \midrule
  T2 + WaveGlow-128  & $3.90 \pm 0.14$ & $4.43 \pm 1.05$ \\
  FS + WaveGlow-128  & $3.82 \pm 0.08$ & $0.15 \pm 0.11$ \\
  AT + WaveGlow-128  & $4.01 \pm 0.09$ & $0.16 \pm 0.06$ \\
  \midrule
  T2 + MelGlow-32   & $ 3.89 \pm 0.15$ & $4.48 \pm 1.08$ \\
  FS + MelGlow-32   & $ 3.83 \pm 0.09$ & $0.16 \pm 0.04$ \\
  AT + MelGlow-32   & $ 4.01 \pm 0.07$ & $0.18 \pm 0.05$ \\
  \bottomrule
  \end{tabular}
  \label{Table-TTS}
  \end{center}
\end{table}

\begin{table}[t]
  \begin{center}
  \caption{The comparison of MOS with 95\% confidence intervals for different residual convolution 
           layers and hidden channels in the kernel predictor. }
  \begin{tabular}{p{3cm}p{1.8cm}<{\centering}p{2cm}<{\centering}}
  \toprule
  \textbf{Method}&\textbf{Parameters}&\textbf{MOS} \\
  \midrule
  MelGlow-KP-32C  & $  10.5 M $ & $ 3.93 \pm 0.06 $  \\
  MelGlow-KP-64C  & $  19.3 M $ & $ 4.11 \pm 0.12 $  \\
  MelGlow-KP-128C & $  38.6 M $ & $ 4.11 \pm 0.09 $  \\
  \midrule
  MelGlow-KP-1L   & $  18.1 M $ & $ 4.05 \pm 0.06 $  \\
  MelGlow-KP-3L   & $  19.3 M $ & $ 4.11 \pm 0.12 $  \\
  MelGlow-KP-5L   & $  20.6 M $ & $ 4.09 \pm 0.11 $  \\
  \bottomrule
  \end{tabular}
  \label{Table-Analysis}
  \end{center}
\end{table}

\section{Conclusion}

In this paper, location-variable convolutions is proposed to 
model the long-term dependency features of waveform 
according to local acoustic features,
where different intervals in the input sequence use 
different convolution kernels to implement convolution operation. 
Meanwhile, it is used to design a efficient neural vocoder, named MelGlow, 
based on the architecture of WaveGlow. 
Experiments on the LJSpeech dataset show that MelGlow achieves better 
performance than WaveGlow in the case of limited model size
due to the improved dependency capture capabilities of location-variable convolutions. 

\section{Acknowledgements}

This paper is supported by National Key Research and 
Development Program of China under grant No. 2017YFB1401202, 
No. 2018YFB1003500 and No. 2018YFB0204400. 
Corresponding author is Jianzong Wang from Ping An Technology (Shenzhen) Co., Ltd.


\bibliographystyle{IEEEbib}
\bibliography{melglow}

\end{document}